\documentclass[journal]{IEEEtran}
\usepackage{amsmath,amsfonts}
\usepackage{algorithmic}
\usepackage{algorithm}
\usepackage{array}
\usepackage[caption=false,font=footnotesize,labelfont=rm,textfont=rm]{subfig}
\usepackage{textcomp}
\usepackage{stfloats}
\usepackage{url}
\usepackage{verbatim}
\usepackage{graphicx}
\usepackage{cite}
\usepackage{hyperref}

\usepackage{multirow}
\usepackage{xcolor,color,framed}

\begin{document}


\title{Federated Agentic AI for Wireless Networks: Fundamentals, Approaches, and Applications}

\author{Lingyi Cai, Yu Zhang, Ruichen Zhang, Yinqiu Liu, Tao Jiang,~\IEEEmembership{Fellow,~IEEE}, Dusit Niyato,~\IEEEmembership{Fellow,~IEEE}, \\ Wei Ni,~\IEEEmembership{Fellow,~IEEE}, and Abbas Jamalipour,~\IEEEmembership{Fellow,~IEEE}

\thanks{Lingyi Cai is with the Research Center of 6G Mobile Communications, School of Cyber Science and Engineering, Huazhong University of Science and Technology, Wuhan, 430074, China, and also with the College of Computing and Data Science, Nanyang Technological University, Singapore (e-mail: lingyicai@hust.edu.cn).}
\thanks{Yu Zhang and Tao Jiang are with the Research Center of 6G Mobile Communications, School of Cyber Science and Engineering, Huazhong University of Science and Technology, Wuhan, 430074, China (e-mail: yuzhang123@hust.edu.cn; Tao.jiang@ieee.org).}
\thanks{Ruichen Zhang, Yinqiu Liu, and Dusit Niyato are with the College of Computing and Data Science, Nanyang Technological University, Singapore (e-mails: ruichen.zhang@ntu.edu.sg; yinqiu001@e.ntu.edu.sg; dniyato@ntu.edu.sg).}
\thanks{Wei Ni is with the School of Engineering, Edith Cowan University, Perth, WA 6027, and the School of Computer Science and Engineering, University of New South Wales (UNSW), Sydney, NSW 2033, Australia (e-mail: Wei.Ni@ieee.org).}
\thanks{Abbas Jamalipour is with the School of Electrical and Computer Engineering,
University of Sydney, Australia, and with the Graduate School of Information
Sciences, Tohoku University, Japan (e-mail: a.jamalipour@ieee.org).}


}



\maketitle

\begin{abstract}
Agentic artificial intelligence (AI) presents a promising pathway toward realizing autonomous and self-improving wireless network services. However, resource-constrained, widely distributed, and data-heterogeneous nature of wireless networks poses significant challenges to existing agentic AI that relies on centralized architectures, leading to high communication overhead, privacy risks, and non-independent and identically distributed (non-IID) data. Federated learning (FL) has the potential to improve the overall loop of agentic AI through collaborative local learning and parameter sharing without exchanging raw data. This paper proposes new federated agentic AI approaches for wireless networks. We first summarize fundamentals of agentic AI and mainstream FL types. Then, we illustrate how each FL type can strengthen a specific component of agentic AI's loop. Moreover, we conduct a case study on using FRL to improve the performance of agentic AI's action decision in low-altitude wireless networks (LAWNs). Finally, we provide a conclusion and discuss future research directions.
\end{abstract}

\section{Introduction} 

Next-generation wireless networks are expected to operate with a high degree of autonomy, proactively adapting to changing network environments and continuously satisfying evolving service demands \cite{11339915,10763424}. In this context, agentic AI has emerged as a promising paradigm that enables intelligent systems to perceive complex scenarios, maintain contextual memories, perform structured reasoning, and execute goal-oriented actions for various wireless applications \cite{11339915}. By integrating perception, memory, reasoning, and action in a closed loop, agentic AI offers a viable pathway toward realizing proactive and self-improving intelligence in future wireless networks \cite{zhao2025agentification}.

Due to the resource-constrained, widespread distribution, and data-heterogeneous nature of wireless networks \cite{10001439,10.1145/3747351}, existing agentic AI, which typically relies on centralized architecture that shares large volumes of raw information, faces several critical challenges. In this case, the centralized or shared-data paradigm incurs substantial communication overhead, while introducing significant privacy risks and regulatory compliance concerns. Moreover, data across AI agents is non-independent and identically distributed (non-IID) due to varying deployment environments and device capabilities, which leads to inconsistent or biased outcomes in reasoning and actions. These limitations undermine the efficiency, privacy, and scalability of agentic AI in distributed wireless network settings.

Federated learning (FL) provides a distributed architecture that has the potential to enable AI agents to collaboratively improve their capabilities across the components of perception, memory, reasoning, and action. By integrating FL into the agentic AI loop, agents perform local learning using their private data and only need to share model parameters or gradients instead of raw data, which effectively preserves data privacy and significantly reduces communication overhead \cite{10001439}. This is particularly critical in wireless environments where bandwidth is constrained and data (e.g., user traffic, channel measurements, and network logs) is often sensitive and subject to regulations. Furthermore, FL can aggregate diverse learning experiences from all agents to synthesize diverse operational experiences and environmental knowledge, thereby improving robustness and generalization of agentic AI across distributed wireless environments \cite{10697226}. In this sense, the integration of FL provides a promising way to overcome key limitations of centralized agentic AI in next-generation wireless systems.



This paper provides a comprehensive overview of how FL can empower agentic AI in wireless networks. We first summarize the fundamentals of agentic AI and mainstream FL types, including federated supervised learning (FSL), federated unsupervised learning (FUL), federated graph learning (FGL), federated generative learning (FGenL), and federated reinforcement learning (FRL). Then, we demonstrate how FL strengthens each component of the agentic AI loop. Specifically, FSL/FUL enables distributed perception from labeled/unlabeled data in wireless environments; FGL facilitates long- and short-term memory extraction and construction without sharing raw relational data; FGenL refines domain-aligned reasoning through fine-tuning a large language model (LLM) and optimizing chain-of-thought (CoT) in a federated manner; and FRL supports tool-augmented action by aggregating decentralized decision-making experiences. Finally, we discuss representative wireless applications of FRL for the action of agentic AI in low-altitude wireless networks (LAWNs) to demonstrate its practical viability and outline future research directions.

The main contributions of this paper are summarized as follows:

\begin{enumerate}
  \item We lay conceptual foundations for the agentic AI and FL paradigm, based on which we uncover the potential of leveraging FL to address limitations of agentic AI in wireless networks.

  \item We propose the novel federated agentic AI approaches for wireless networks, where FSL/FUL, FGL, FGenL, and FRL are utilized to empower the perception, memory, reasoning, and action of agentic AI, respectively.

  \item We provide a case study of using FRL to empower agents' actions to defend jamming attacks in LAWNs, which shows that FRL can achieve more scalable, efficient, and robust defense actions compared to the centralized action decision form.

\end{enumerate}

\section{Fundamentals of Agentic AI and FL } 

In this section, we establish the foundational concepts necessary for understanding the integration of FL with agentic AI in wireless networks. 

\subsection{Fundamentals of Agentic AI}

Agentic AI represents a paradigm shift from the passive and single-shot inference models (such as single-turn LLMs) toward autonomous intelligent systems capable of continuous improvement \cite{11339915}. Agentic AI can organize more AI agents into closed loops that can proactively decompose objectives, perceive changing environments, update persistent memories, perform structured multi-step reasoning, coordinate tool-based actions, and refine strategies for evolutionary improvement \cite{zhao2025agentification}. This paradigm shift is compelling for wireless networks due to the characteristic of high dynamics, task heterogeneity, and resource constraints. Specifically, the complete loop of agentic AI for wireless networks comprises four core components:

\subsubsection{Perception} The perception layer is responsible for gathering and pre-processing heterogeneous data from the environment. Next-generation wireless networks include multimodal data, such as  visual feeds, network state information, textual logs, and sensor data, that go beyond traditional radio-frequency (RF) signals. These data can be processed by specific models (e.g., CNNs for images and signal processing modules for RF) to perform initial feature extraction and fusion. Then, the processed data is contextualized and normalized to provide a unified situational awareness crucial for wireless network tasks.

\subsubsection{Memory} The memory component endows the agent with long-term knowledge and short-term context about wireless networks. The former is stored in vector databases or knowledge graphs (KGs), including domain-specific expertise such as network topology history and communication standards. Agents can utilize this knowledge base to make decisions based on factual and up-to-date information. Short-term memory maintains the immediate context of an ongoing task or interaction, such as recent observations, intermediate results, or dialogue history.

\subsubsection{Reasoning} The reasoning component enables the agents to process the perceived context and retrieved knowledge for decision-making in wireless networks. At the fine-tuning stage, the reasoning capability can be strengthened through Supervised Fine-Tuning (SFT) and Reinforcement Learning from Human Feedback (RLHF), which align intermediate reasoning steps and final decisions with wireless-relevant objectives. At the inference phase, Chain-of-Thought (CoT) prompting and self-consistency decoding can elicit structured step-by-step solutions under dynamic network conditions. 

\subsubsection{Action} The action layer translates reasoning results into concrete actuation in the wireless environment. Action execution is guided by strategies in preceding reasoning steps, where agents can call predefined or self-generated tools for wireless networks, such as API calls, simulation platforms, or control interfaces. Actions may make changes to the network state, generate new observations, or trigger updates in memory. This feedback is incorporated into subsequent loops of agentic AI for iteration.

\subsection{Fundamentals of FL} 

FL paradigms enable end devices to use private data to train local models and upload them to a coordinating server to aggregate a global model \cite{10001439,10428063}. From the viewpoint of training objectives, mainstream FL paradigms in wireless networks can be broadly categorized into four types:

\emph{ 1) Federated supervised learning (FSL)} primarily addresses classification and regression tasks with labeled data. The server only collects model parameters and aggregates a global model using algorithms such as FedAvg \cite{10001439}. \emph{Federated unsupervised learning (FUL)} addresses scenarios where labeled data is scarce or unavailable, focusing on tasks such as clustering and dimensionality reduction. Clients locally extract features and then exchange model parameters or representation statistics with the server \cite{10697226}.

\emph{2)  Federated graph learning (FGL)} integrates FL with graph neural networks (GNNs) to handle graph-structured data in distributed settings. This paradigm addresses the unique challenge of training GNN models when graph data is partitioned across multiple clients with privacy constraints \cite{10428063}, especially for tasks such as KG completion, graph classification, and social recommendation.

\emph{3) Federated generative learning (FGenL)} integrates FL with generative AI technologies to enable training or fine-tuning of generative models (e.g., LLMs, generative adversarial networks, and diffusion models), representation learning, and content generation in a privacy-preserving manner, where distributed clients can refine local generative models based on domain-specific data for global model aggregation \cite{10398264}.

\emph{4) Federated reinforcement learning (FRL)} combines reinforcement learning (RL) with FL to collaboratively optimize decision-making policies. Multiple distributed RL agents interact with local environments, update their policies based on local rewards, and periodically share local policies for global aggregation while keeping their local observations private \cite{10691666}.

\subsection{Motivation of Integrating FL into Agentic AI}

While agentic AI promises autonomously and continuously improving intelligence for wireless networks, its realization is challenged by data privacy concerns, network heterogeneity, and resource constraints. FL offers promising opportunities to address these challenges for the execution of perception, memory, reasoning, and action of agentic AI. The core advantage of this integration lies in preserving data privacy, enabling adaptation to network dynamics, and leveraging diverse local experiences to enhance the complete loop of agentic AI for next-generation wireless networks.

\begin{figure*}[t]
\centering
\includegraphics[width=1\linewidth]{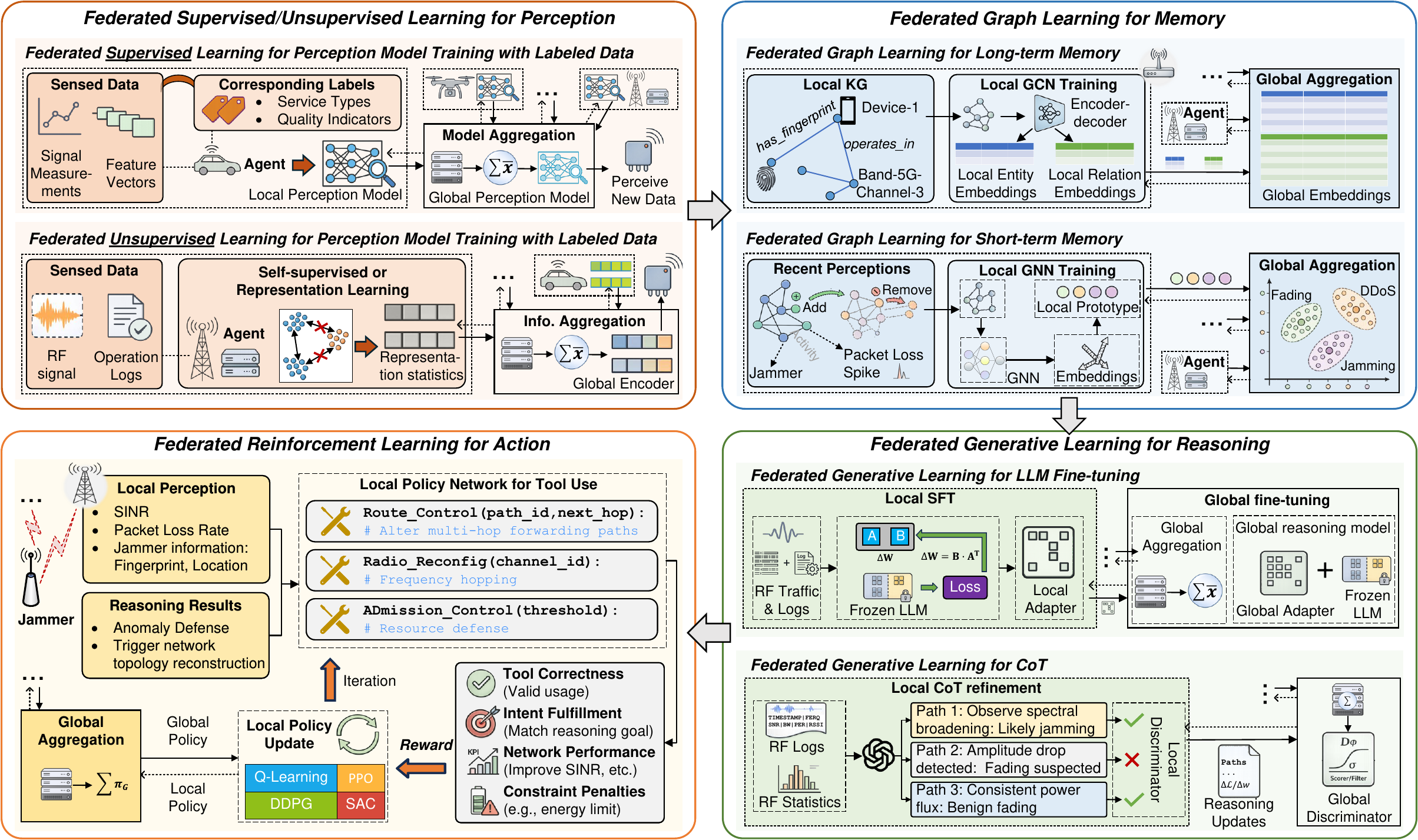}
\caption{Overview of a federated agentic learning architecture for wireless networks. FSL and FUL enables privacy-preserving perception from decentralized RF measurements and logs. FGL constructs short- and long-term memory from distributed knowledge graphs. FGenL refines domain-specific reasoning, including LLM fine-tuning and CoT optimization. FRL drives tool-augmented actions through collaboratively learned policies. These components form a unified perception–memory–reasoning–action loop for RF anomaly detection and anti-jamming in wireless environments.}
\label{fedagenticaifram}
\end{figure*}

\section{Federated Agentic AI Approaches for Wireless Networks} %

In this section, we present a detailed framework for integrating various FL paradigms into each component of the Agentic AI loop, as shown in Fig. \ref{fedagenticaifram}. First, we examine how FSL and FUL enable privacy-preserving perception from distributed labeled and unlabeled data, including applications such as RF fingerprint identification. We then explore how FGL facilitates the collaborative construction and updating of memory representations (e.g., knowledge graphs) without sharing raw and privacy-sensitive data. Following this, we describe how FGenL enhances reasoning capabilities through the federated fine-tuning of LLMs using low-rank adapters and the refinement of CoT logic by aggregating high-quality reasoning chains. Finally, we discuss how FRL supports adaptive and tool-augmented action policies by federating local policy updates based on environmental feedback and reward signals. 


\subsection{Federated Supervised/Unsupervised Learning for Perception}

The perception component of agentic AI is required to continuously transform complex environmental signals into actionable state representations as the foundation for reasoning and decision making. However, raw perception data are often isolated across multiple distributed nodes in wireless networks due to privacy concerns, which imposes significant limitations on traditional centralized learning methods for perception model training. The FSL and FUL enable agents to collaboratively build a global perception model without sharing raw data, thereby equipping them with comprehensive, accurate, and adaptive environmental understanding capabilities.

For tasks where labeled data is available, FSL enables agents to train a local perception model, such as a signal classification model, using the sensed data (e.g., traffic feature vectors or radio signal measurements) together with corresponding labels (e.g., network service types or channel quality indicators). Local models are uploaded to a central server to aggregate a global perception model, which is distributed back to the agents to accurately perceive new data. Taking RF fingerprint identification as an example \cite{10001439}, distributed agents equipped with receivers can train local CNN models on locally labeled In-phase/Quadrature samples. These local models are periodically uploaded to form a global model as a classifier, which captures invariant device-specific signatures across heterogeneous sensing sites. Through such supervised federation, agents deployed under heterogeneous channel conditions can use the global model to verify the identity of devices in the perception process.

When large amounts of perception data lack labels, rather than direct classification in FSL, FUL enables clients to leverage local unlabeled perception data (e.g., unannotated RF signal segments and device operation logs) for self-supervised or representation learning. Each agent learns to pull similar signal patterns closer and push dissimilar ones farther apart in the embedding space. Representation statistics (such as distribution moments of embedding vectors) are exchanged among agents, and the server aggregates this information to form a global encoder capable of extracting task-specific features. The study in \cite{10697226} utilizes FUL for perceptual feature extraction in RF fingerprint identification. By adopting the contrastive learning method, distributed receivers locally train feature extractors on unlabeled RF signals, where local parameters are aggregated at a server to form a global feature encoder. Thus, distributed agents can collaboratively perceive feature representations of invariant RF fingerprints for device identification through FUL when a continuous influx of novel and unannotated signals occurs in dynamic wireless environments.

\subsection{Federated Graph Learning for Memory}

The memory preserves persistent knowledge and contextual signals of wireless environments through the perception component. However, knowledge relevant to wireless networks, such as operation patterns and historical connectivity, is fragmented across numerous devices. Though this knowledge can be constructed in a graph form for the memory of agentic AI, graph partitions across agents have privacy-sensitive relationships that cannot be directly managed in a centralized manner. FGL enables agents to collaboratively build and update memory representations without sharing raw graph data, which can enable agents to be equipped with comprehensive and up-to-date knowledge of wireless environments for use by downstream reasoning modules.


The long-term memory of agentic AI can be enhanced by collaboratively learning historical knowledge for entities and relations in distributed KGs. However, perceptual knowledge is fragmented across distributed agents due to privacy constraints and spatial heterogeneity, making centralized KG construction infeasible in practical wireless deployments. FGL has the ability to construct and continuously maintain long-term memory by learning embeddings from distributed KGs. These embeddings contain raw features (e.g., power and type) of the node and encode the node’s relational context and interaction history in the graph. For example, KG entities (e.g., ``Device-1", ``Band-5G-Channel-3") and relations (e.g., ``has\_fingerprint", ``operates\_in") perceived via FSL encode the association between a device and its characteristic RF features, while similarity clusters and temporal co-occurrence patterns perceived via FUL are formalized as relations in a local KG, such as ``similar\_to" or ``occurs\_before". The structural embeddings of these subgraphs can be learned by agents employing a local Graph Convolutional Network (GCN)–based encoder–decoder \cite{10.1145/3747351}. The local entity and relation embeddings are uploaded to the server for aggregating global embedding, which builds a persistent long-term knowledge foundation that accumulates domain expertise over time.

Short-term memory can be realized by dynamically organizing recent perception outputs into a memory graph, where nodes represent entities (e.g., devices and RF signal segments), and edges encode interactions or correlations \cite{ijcai2025p2}. The graph structure is dynamically updated by adding new nodes and edges while removing outdated information. For instance, a jammer perceived by FUL with a feature vector \text{[band, location, jamming period]} is a node, while an edge (``Jammer” $\rightarrow$ ``Packet Loss Spike") indicates that a packet loss peak is observed in the network after the jammer’s activity. Once observations are represented in the graph form, federated GNNs can be applied for processing and representation learning, where each agent can maintain a local memory model aligned with its private context and benefit from globally shared knowledge \cite{pmlr-v260-xie25b}. Specifically, each agent trains a local GNN model to process the memory graph to obtain embeddings, from which local prototypes (i.e., the centroid of all embeddings) are computed. Agents upload local prototypes to a central server, where contrastive learning can be used to cluster all local prototypes into different types (e.g., benign channel fading, distributed denial of service (DDoS) pattern, and jamming pattern) to construct global prototypes. The updated global prototypes are distributed back to guide agents to further train their local GNN models to enhance short-term memory.

\subsection{Federated Generative Learning for Reasoning}

Agentic AI equipped with conventional pre-trained LLMs possesses broad general knowledge, but may lack the deep understanding of the complex scenarios and dynamic tasks of wireless networks during reasoning. FGenL is promising for coordinating multiple agents to use private data to adapt the model to a specific domain and to enhance reasoning capabilities without compromising privacy.  By utilizing the perception and memory that provide factual and relational context, the fine-tuned LLMs and CoT can reason more accurately and explainably. The following discussion provides insights into applying FGenL to fine-tune LLMs and enhance CoT reasoning.

A key application of FGenL is the collaborative refinement of LLMs for reasoning in agentic AI. The standard pre-trained LLMs with general knowledge lack the specialized and domain-specific understanding required for complex wireless tasks. Taking RF anomaly detection applications as an example, the work in \cite{11059535} has shown that fine-tuned LLMs can effectively identify anomalous network behaviors by reasoning over textualized RF traffic and log data. However, in practical wireless deployments, centralized fine-tuning of LLMs has limitations since RF measurements and operational logs are decentralized and privacy-sensitive. FGenL has the potential to provide a framework to enable each agent (e.g., on a base station or a network controller) to leverage its local data to perform SFT on a local copy of the LLM \cite{10763424}. Specifically, each agent injects a trainable low-rank adapter (LoRA) module into a frozen pre-trained LLM. During the local fine-tuning phase, only the parameters of this adapter module are updated via backpropagation on the local data, thereby computing local gradients. Subsequently, these local gradients are transmitted to a central server to update the global low-rank adapter module. Through multiple rounds of local fine-tuning, the global adapter effectively integrates knowledge from diverse RF conditions, traffic dynamics, and attack behaviors. Finally, the global reasoning model can be constructed by integrating the global adapter with the original frozen LLM. 


Furthermore, strengthening the CoT capability is critical for enhancing the reasoning functions of agentic AI. Traditional CoT approaches primarily rely on centrally maintained LLMs guided by manually engineered CoT prompts \cite{wang2025chain}, which assume that task-relevant knowledge can be aggregated at a single location. By utilizing FGenL, distributed agents can construct a global CoT logic that internalizes diverse operational reasoning patterns, thereby producing high-quality and domain-aligned CoT reasoning chains for wireless tasks. For example, each agent employs a local generative reasoning model to autonomously generate multiple candidate CoT reasoning paths that explain observed anomalies in RF, such as distinguishing jamming-induced distortions from benign channel fading. A local discriminator is then trained to evaluate and retain high-quality reasoning chains based on their consistency with local RF statistics and detection outcomes. Only these informative reasoning updates are uploaded to the server for federated aggregation to construct a global discriminator, which is subsequently redistributed to guide inference-time CoT selection at each agent for RF anomaly detection.

\subsection{Federated Reinforcement Learning for Action}

In wireless networks, actions of agentic AI are coupled with dynamic environments and operational constraints, making centralized decision-making impractical due to latency, bandwidth overhead, and privacy. By introducing FRL into the action layer, each agent, such as base stations and UAVs, can autonomously execute actions based on private local observations \cite{10.1145/3746252.3761391}. Through continuous interaction, agents explore diverse action strategies, especially tool-augmented actions such as API invocations, simulator queries, and control-interface calls. These local experiences can be aggregated into a global action policy that captures collective operational knowledge across the network. Thus, FRL enables the action layer of agentic AI to function as a self-improving system.

Taking the RF anomaly scenario as an example, each AI agent operates as a local FRL agent, which utilizes local perception and reasoning results as inputs. The local perception includes communication quality indicators (e.g., signal-to-interference-plus-noise ratio (SINR) and packet loss rate) and information related to jammers (e.g., fingerprints and locations). The reasoning results convey intents such as anomaly defense or triggering network topology reconstruction. The FRL agents are expected to output a sequence of tool invocations, which specify each concrete tool along with its parameters to translate the reasoning intent into executable control operations. For example, the sequence may include multiple tools such as a routing-control interface that allows the agent to deliberately alter multi-hop forwarding paths in response to link-level attacks, a radio reconfiguration interface for frequency hopping, and resource-defense interfaces for admission control. The local FRL agent evaluates its action decisions using a composite reward signal that reflects the correctness of tool invocations and the resulting network performance. The reward design can incorporate multiple aspects and is not limited to motivating the tool sequence to fulfill the received reasoning results and penalizing actions that violate operational constraints such as energy limits.

The reward integrates real-time feedback from the environment. For instance, an FRL agent at a base station executes an action (e.g., adjusting the radio reconfiguration interface) that changes the local network state such as SINR. This change can be observed by the agent and generate new observations and local rewards, which are used to train the local policy iteratively. The policy learning process can follow the standard policy-gradient-based RL algorithm such as Proximal Policy Optimization (PPO), Deep Deterministic Policy Gradient (DDPG), and Twin Delayed DDPG (TD3) \cite{10691666}. The local policy updates are periodically uploaded to a central server such as a ground control station. The server aggregates these updates using a federated averaging mechanism to obtain a global policy, which is then fed back to all agents for deployment in the next training round.


\subsection{Lesson Learned}
This section demonstrates the successful integration of FL with agentic AI in a modular and synergistic manner for wireless networks, where various FL paradigms are strategically empowered to specific components—perception, memory, reasoning, and action. This design transitions from centralized data processing to distributed knowledge fusion, enabling the emergence of efficient, scalable, and privacy-preserving agentic AI. Additionally, the proposed federated agentic AI faces potential practical challenges that deserve attention, such as inter-agent synchronization and orchestration overhead, which are critical issues for future practical deployments.

\section{Case Study: FRL for Action of Agentic AI in Jamming Defense in LAWNs}

\begin{figure}[t]
\centering
\includegraphics[width=1.0\linewidth]{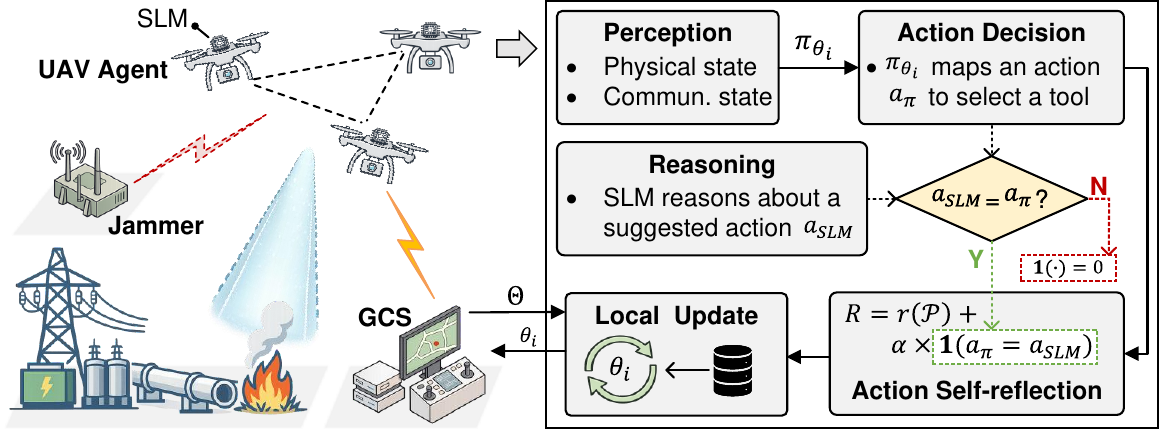}
\caption{(Left): The case-study scenario of a jamming-resilient UAV swarm in the LAWNs. (Right): The overall workflow of the FRL for actions of UAV agents, illustrating perception of physical and communication states, SLM-assisted reasoning, tool-based action decision and self-reflection, local policy update, and federated policy aggregation.} 
\label{casestudyfig}
\end{figure}

\subsection{System Overview}

We consider a jamming-resilient UAV swarm system deployed for emergency response over critical infrastructure in the LAWNs, as shown in Fig. \ref{casestudyfig}. The system consists of a set of cooperative UAVs and a trusted ground control station (GCS). One UAV is designated as the logical leader, responsible for receiving high-level mission commands from the GCS and disseminating control and coordination information to follower UAVs. The swarm is vulnerable to jamming attacks due to the broadcast nature of wireless communication in low-altitude airspace. The jammer transmits interference signals to jam the communication channels. Each UAV is augmented with a pre-trained small language model (SLM) (e.g., TinyLlama), serving as an AI agent for autonomous reasoning in the jamming defense. The SLM can be fine-tuned by the GCS using communication signal monitoring data and RF logs. For instance, an input prompt to the SLM encodes the local obseravtion of UAV agent with the executed jamming defense action, while the target output comprises whether an attack has occurred and an assessment of the action taken. The fine-tuned SLM can provide knowledge-based reasoning as references and interpretability for decision-making in jamming defense.

The swarm operates under a set of constraints, including prescribed UAV cruising speed, heading angle, inter-UAV safety distance, and allowable communication frequency ranges. The system’s objective $\mathcal{P}$ is to jointly minimize the number of communication links successfully attacked by the jammer and the cost incurred by the UAV agents when invoking defense tools. We devise new FRL to address this multi-objective optimization problem, providing a distributed policy learning framework in which each UAV agent operates as an independent learner and optimizes its action decisions based on private local observations and experiences. Moreover, FRL enables collaborative decision update within the swarm without sharing raw observation data. To the best of our knowledge, this is the first advanced FRL paradigm designed specially for this challenging scenario.

\subsection{Workflow of FRL for Action of Agentic AI in Jamming Defense}

Overall, FRL for the action component of agentic AI is mainly implemented through local action policy learning at the UAV agents and global action policy aggregation at the GCS. Each UAV agent uses a policy-gradient algorithm to optimize its defense tool selection decisions based on local observations, the suggested actions reasoned about by the SLM, the actions taken by the policy network, and the received rewards. After multiple training episodes, the agents upload their model parameters to the GCS for global action policy aggregation, enabling collaborative decision updates.

\subsubsection{\bf Local Action Policy $\theta_i$ Learning} Each UAV $i$ acts as an independent AI agent, which is equipped with an SLM for decision-making assistant, and maintains a local policy network for action selection. The action of the UAV agent is to select tools to defend against jamming attacks, where the available defense tools are three predefined interfaces built into the UAV's communication and control stack, i.e., $Role\_Shuffling()$, $Topology\_Reconfiguration()$, and $Frequency\_Hopping()$. The $Role\_Shuffling()$ tool reassigns the logical leader role to a new UAV when the link between the current leader and the GCS is persistently targeted by jamming attacks. The $Topology\_Reconfiguration()$ tool updates local routing tables to reconfigure the communication topology to bypass jammed links. The $Frequency\_Hopping()$ tool proactively changes the communication frequency channels of the UAVs and the GCS to avoid persistent jamming.

\begin{itemize}

\item {Perception:} The UAV agent can perceive its physical state (e.g., position, velocity, and heading). Meanwhile, it can perceive the current communication state, including its role in the communication topology (i.e., leader or follower), current frequency, and whether its communication links are connected.

\item {Reasoning:} The SLM mounted on a UAV reasons about suggested actions (i.e., tool usage) based on its memory and knowledge base. For example, taking the local observation that the physical state is normal, this UAV is not the leader, and the communication link is disconnected as input, the SLM can reason that the jamming attack is targeting the link between this follower UAV and the leader UAV. Accordingly, the suggested action $a_{SLM}$ is to select the $Topology\_Reconfiguration()$ tool.

\item {Action Decision:} The UAV agent uses the local policy network $\pi_{\theta_i}$ to map the current state to an action $a_{\pi}$ to select a tool for defending against the jamming attack.

\item {Action Self-reflection:} We propose to provide the UAV agent with a carefully designed reward signal $R = r(\mathcal{P}) + \alpha \times \textbf{1}(a_{\pi}=a_{SLM})$ to guide its self-reflection on action selection, where the reward signal consists of two components. The first component $r(\mathcal{P})$ of the reward signal is a weighted sum of optimization terms corresponding to the system objective $\mathcal{P}$. This follows the classical RL reward design to guide the agent’s actions toward optimizing the objective. The second component is related to the SLM’s reasoning results. If the action $a_{\pi}$ taken by the UAV agent is consistent with the suggested action $a_{SLM}$, i.e., $a_{\pi}=a_{SLM}$, an additional reward $\textbf{1} (a_{\pi}=a_{SLM})$ is provided; otherwise, this term equals zero. In addition, $\alpha$ denotes a decay weight. It guides the agent to avoid random exploration in a large search space in the early training stage. In the later training stage, as the agent can make decisions based on accumulated experience, $\alpha$ is gradually decayed to reduce the intervention of the SLM.

\item { Local Policy Update:} The UAV agent stores its learning experiences in a local experience replay buffer and updates the parameters $\theta_i$ of the local policy network using a policy-gradient algorithm.

\end{itemize}


\subsubsection{\bf Global Action Policy $\Theta$ Aggregation} Each UAV agent uploads its local policy parameters $\theta_i$ to the GCS to perform federated averaging (i.e., $\Theta = \sum_{i=1}^{N} \frac{1}{N}\,\theta_i$) to aggregate the global policy, where $N$ is the total number of UAVs in the swarm. Then, the global policy is fed back to the UAV agents to update their local policies. This iterative federated process enables the UAV swarm to eventually learn the optimal action policy.

\subsection{Performance Evaluation}

\begin{figure}[t]
\centering
\includegraphics[width=1.0\linewidth]{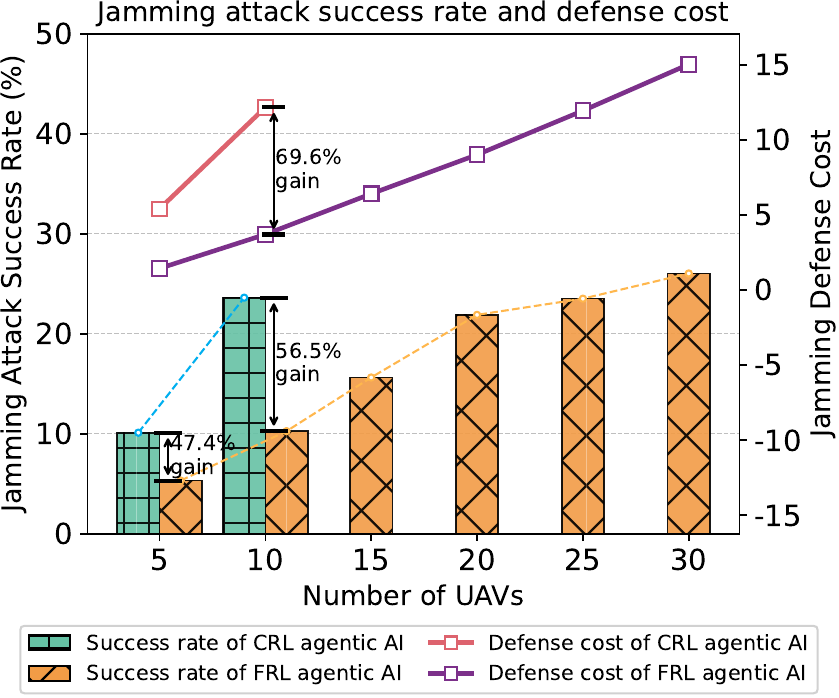}
\caption{Jamming attack success rate and defense cost of FRL agentic AI and CRL agentic AI under different UAV swarm sizes. As the swarm size increases, the centralized coordination overhead at the GCS grows rapidly, making CRL difficult to scale beyond N=10; thus, results of CRL for larger swarm sizes are not included.} 
\label{costratemagafig}
\end{figure}

In the simulation settings, we consider a $1 \times 1~\text{km}^2$ low-altitude monitoring area with a GCS located at the center. The system consists of a UAV swarm operating in a circular formation with a radius of 200 m and a fixed altitude of 100 m. To demonstrate the effectiveness of our proposed FRL for agentic AI (denoted as FRL agentic AI), we adopt the centralized RL \cite{9563249} for agentic AI (denoted as CRL agentic AI) as the baseline for comparative experiments, thereby evaluating the UAV agents’ defense tool selection (i.e., actions).

Fig. \ref{costratemagafig} jointly illustrates the jamming defense cost and attack success rate of FRL agentic AI and CRL agentic AI under different UAV swarm sizes. The performance of the proposed FRL agentic AI consistently outperforms the CRL agentic AI. In particular, when the UAV number is $N=10$, FRL agentic AI achieves a defense cost reduction of approximately 69.6\% relative to CRL agentic AI, while simultaneously reducing the attack success rate by about 56.5\%. This performance advantage can be attributed to the fact that CRL suffers from a rapidly expanded action space induced by centralized multi-agent decision-making, which leads to suboptimal and high-cost defense policies. In contrast, FRL decomposes the centralized decision process into distributed local policies with federated coordination, thereby enabling more scalable, cost-efficient, and robust jamming defense.

\section{Conclusion}

In this paper, we have explored federated agentic AI approaches in wireless networks by systematically integrating different types of FL into the perception–memory–reasoning–action loop of agentic AI. A case study on FRL-based anti-jamming action in the LAWNs shows that FRL for action reduces defense cost by 69.6\% and lowers the attack success rate by 56.5\% compared to a centralized action decision. However, the trustworthiness of the federated process deserves sufficient attention, as malicious attackers may launch poisoning attacks or inference attacks. Future research can further focus on trustworthy federated coordination such as Byzantine-resilient aggregation and privacy-enhancing model sharing, thereby realizing more reliable agentic AI in next-generation wireless networks.

\bibliographystyle{IEEEtran}
\bibliography{ref}

\vfill

\end{document}